\documentclass[conference,a4paper]{IEEEtran}
\usepackage[left=1in, right=0.7in, bottom=1.4in, top=0.76in]{geometry}
\IEEEoverridecommandlockouts

\usepackage{cite}
\usepackage{amsmath,amssymb,amsfonts}
\usepackage{algorithmic}
\usepackage{graphicx}
\usepackage{textcomp}
\usepackage{xcolor}
\usepackage{fancyhdr}
\usepackage{url}
\def\BibTeX{{\rm B\kern-.05em{\sc i\kern-.025em b}\kern-.08em
    T\kern-.1667em\lower.7ex\hbox{E}\kern-.125emX}}
\begin{document}
\makeatletter

\title{Comparative Analysis of Clustering Methods for Power Delay Profile in MMW Bands and In-Vehicle Scenarios \\}

\author{\IEEEauthorblockN{ Radek Zavorka\IEEEauthorrefmark{1}, Ales Prokes\IEEEauthorrefmark{1}, Josef Vychodil \IEEEauthorrefmark{1}, Tomas Mikulasek\\ \IEEEauthorrefmark{1}, Petr Horky \IEEEauthorrefmark{1},Christoph Mecklenbräuker \IEEEauthorrefmark{2}, Aniruddha Chandra\\ \IEEEauthorrefmark{3}, Jan Marcin Kelner \IEEEauthorrefmark{4}, Cezary Henryk Ziółkowski
 \IEEEauthorrefmark{4}}
\IEEEauthorblockA{\IEEEauthorrefmark{1}Department of Radio Electronics, Brno University of Technology, Brno, Czech Republic}
\IEEEauthorblockA{\IEEEauthorrefmark{2}Institute of Telecommunications, TU Wien, Vienna, Austria}
\IEEEauthorblockA{\IEEEauthorrefmark{3}National Institute of Technology, Durgapur, India}
\IEEEauthorblockA{\IEEEauthorrefmark{4}
Institute of Communications Systems, Faculty of Electronics,\\ Military University of Technology, 00-908 Warsaw, Poland}
e-mail: xzavor03@vutbr.cz}

\fancypagestyle{firstpage}
{
    \fancyhead[C]{The paper has been presented at the 2023 IEEE Conference on Antenna Measurements and Applications (CAMA), Genoa, Italy, November 15-17, 2023, \url{https://doi.org/10.1109/CAMA57522.2023.10352767}}    
    \fancyfoot[C]{This research was funded in part by the National Science Center (NCN), Poland, grant no. 2021/43/I/ST7/03294 (MubaMilWave). For this purpose of Open Access, the author has applied a CC-BY public copyright license to any Author Accepted Manuscript (AAM) version arising from this submission.}
}

\maketitle

\thispagestyle{firstpage}

\begin{abstract}
The spatial statistics of radio wave propagation in specific environments and scenarios, as well as being able to recognize important signal components, are prerequisites for dependable connectivity. There are several reasons why in-vehicle communication is unique, including safety considerations and vehicle-to-vehicle/infrastructure communication.

The paper examines the characteristics of clustering power delay profiles to investigate in-vehicle communication. It has been demonstrated that the Saleh-Valenzuela channel model can also be adapted for in-vehicle communication, and that the signal is received in clusters with exponential decay. A measurement campaign was conducted, capturing the power delay profile inside the vehicle cabin, and the reweighted $\ell_1$ minimization method was compared with the traditional k-means clustering techniques.

\end{abstract}


\section{Introduction}
Channel modeling is a significant part of almost all communication systems. We need to understand how the signal propagates from point A to point B in a certain environment if reliable communication is required or new standards or frequency bands should be implemented \cite{molish}.

A prominent stochastic model for indoor multipath propagation is the Saleh-Valenzuela (SV) model \cite{SV}. According to the SV model, the tap gain coefficients of the channel impulse response (CIR) between transmitter (TX) and receiver (RX) and also in individual clusters have exponential decrease. A ragged environment, propagation losses, and reflection potentially causes clusters of multi path components (MPC) from the received signal. The model was proposed for the indoor environment of mid-size rooms \cite{SV_molish}, but we show that the model is applicable to in-vehicle communication.

Given that modern car wire harnesses often weigh more than several tens of kilograms \cite{save_wire}, millimeter-wave (MMW) communication can also be used as a wireless bus to support non-critical safety functions. It is essential to systematically analyze the spatial signal distribution inside the passenger cabin in various conditions to ensure dependable connectivity.

In \cite{spread_loss}, researchers focused on examining and comparing delay spread and path loss at UWB 3–11 GHz and MMW 55–65 GHz. In the work \cite{aircraft}, the in-vehicular wireless 60 GHz channel for the aircraft cabin is described. The measurement campaign outlined in the paper \cite{vibration} was used to analyze the effects of driving-related vibrations and twisting on CIR and delay-Doppler spread (DDS) inside the car cabin.

Data clustering proves invaluable in the identification of outliers and determination of signal components, particularly within complex datasets. A wide array of methods, ranging from straightforward to highly intricate, have been harnessed for this purpose. Among these, the unsupervised k-means algorithm has gained notable popularity \cite{kmeans1, kmeans2}. Nevertheless, a critical aspect is specifying the optimal number of clusters that the algorithm should delineate.

An intriguing alternative lies in the Sparsity-based clustering technique, which adeptly segregates the Power Delay Profile (PDP) into clusters without necessitating a predetermined cluster count. Detailed discussions of this method can be found in the works \cite{rwl1, rwl1.1}, grounded in the principles of reweighted $\ell_1$ minimization.


\subsection{Contribution of the Paper}
This paper is focused on analyzing clustering methods for measured millimeter waves in the delay domain in frequency band 55-65 GHz and intra-vehicle scenario. The benefits of paper can be divided into following areas:

\begin{itemize}
    \item comparison of k-means method and Sparsity based clustering based on reweighted $\ell_1$ minimization
    \item analyzing of PDP for in-vehicular communication from measurement campaign by the above mentioned algorithms
    \item validation of the SV model for and intra-vehicle communication at MMW band
\end{itemize}
The paper is organized as follows:

The standard SV channel model is presented in section \ref{section_SV}. The following sections go through clustering techniques, ranging from basic k-means in section \ref{section_kmeans} to more intricate Sparsity based clustering- Reweighted $\ell_1$ minimization in section \ref{section_rwl}. Section \ref{section_exp_me} of the paper regarding experimental measurement provides details on the measurement setup and the application of the aforementioned clustering algorithms to the observed data. Conclusion section sums up the paper.

\section{Saleh-Valenzuela model} \label{section_SV}
Channel models for indoor communication consist of a large number of multipath components caused by reflections, scattering, and diffraction. Given below is the general, complex, low-pass channel impulse response:

\begin{equation}
    \label{rce:CIR_h}
    h(\tau) = \sum_{n=1}^{N} \beta_{n} \exp(j\phi_{n}) \delta(\tau-\tau_{n}),
\end{equation}
where $\beta_{n}$ and $\phi_{n}$ are the amplitude gain and phase of the $n$th path, respectively. $N$ is the total number of propagation paths. $\tau_n$ is the propagation delay (arrival time) of the $n$th path. $\delta$(.) is the Dirac delta function. \cite{rwl1, SV, proakis2}

CIR according to SV model \cite{SV} describing indoor channel is expressed by

\begin{equation}
    \label{rce:CIR_by_SV}
    h(\tau) = \sum_{l=0}^{\infty} \sum_{k=0}^{\infty} \beta_{kl} \exp(j\theta_{kl}) \delta(t-T_{l}-\tau_{kl}),
\end{equation}
where $\beta_{kl}$ and $\theta_{kl}$ are the amplitude gain and phase of the $k$th path within the $l$th cluster, respectively. $T_l$ is delay of $l$th cluster arrival. $\tau_{kl}$ is the excess arrival delay of the $k$th path within the $l$th cluster. \cite{SV}

\begin{figure}[!h]
  \begin{center}
    \includegraphics[scale=1.8]{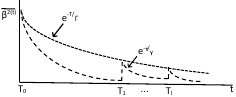}
  \end{center}
  \vspace{-15pt}
  \caption{SV channel model - Exponentially decaying ray and cluster average powers \cite{SV}}
  \label{SV_model1}
\end{figure}

\begin{figure}[!h]
  \begin{center}
    \includegraphics[scale=1.8]{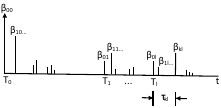}
  \end{center}
  \vspace{-10pt}
  \caption{SV channel model - A realization of the impulse response \cite{SV}}
  \label{SV_model2}
\end{figure}

The most popular approach based on paper \cite{SV} assumes a monotonically decreasing received power with the function of $T_t$ and $\tau_{kl}$ and average power of $\beta_{kl}$ is described by equation

\begin{equation}
    \label{rce:SV_model}
    \overline{\beta_{kl}^2} = \overline{\beta_{0,0}^2} \exp(-\frac{T_l}{\Gamma})\exp(-\frac{\tau_{kl}}{\gamma}),
\end{equation}
where $\overline{\beta_{0,0}^2}$ is the average power gain of the first path of the first cluster, and $\Gamma$ and $\gamma$ are power-delay time constants for the clusters and the paths, respectively \cite{SV}. The Figs. \ref{SV_model1} and \ref{SV_model2} represent the described model.

\section{Channel description}
In-vehicle radio channel measurements are typically assumed to exhibit time-invariant characteristics. So, in order to properly define the radio channel, the CTF is as follows:
\begin{equation}
\label{rce:CTF}
H(k)=s_{21}(k),
\end{equation}
where $k$ is the measurement index that can be linked to a specific frequency. We transform the CTF into a CIR using an inverse Fourier transform in accordance with:
\begin{equation}
\label{rce:CIR}
h(n)=\sum_{k=0}^{N-1} H(k)\mathrm{e}^{jkn2\pi/N,}
\end{equation}
where, in the delay domain, $n$ denotes a discrete time.

The PDP is then determined as the average of a several realizations in accordance with:
\begin{equation}
\label{rce:E}
P(n)= \mathsf{E}\{|h(n)|^2\}.
\end{equation}

\section{Methods of radio signal clustering}
In order to cluster radio signals, a variety of methods are available. From fundamental machine learning techniques like k-means \cite{kmeans3} and enhanced \mbox{k-PowerMeans} \cite{fialova_clustering} up to sparsity-based techniques derived from reweighted $\ell_1$ minimization \cite{rwl1, rwl1.1}.

\begin{figure}[!h]
  \begin{center}
    \includegraphics[scale=0.48]{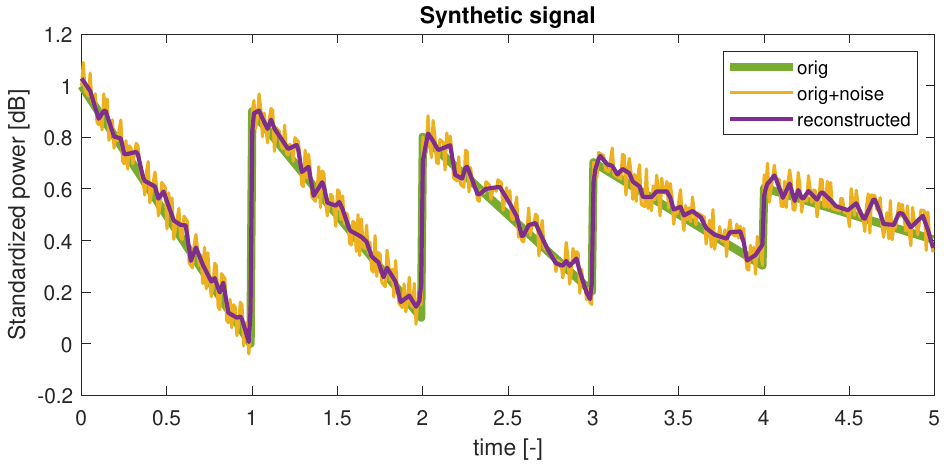}
  \end{center}
  \vspace{-10pt}
  \caption{Synthetic signal for verification of clustering algorithms}
  \label{fig:synthetic}
\end{figure}

Synthetic signals were generated in MATLAB for validating the clustering approaches. Figure \ref{fig:synthetic} shows an exemplary realization of a synthetic signal. There is a transmit signal and received signal that have been impacted by noise, and a signal that has been reconstructed via reweighted $\ell_1$ minimization (explained below).

\section{K-means clustering algorithm} \label{section_kmeans}
Most likely one of the most well-known unsupervised clustering algorithms, applied in many fields \cite{kmeans1, kmeans3}. The number of expected clusters must be specified for proper working. The basic idea behind the algorithm is to find a predetermined number of centroids in the space among points. Measured data from the graph is separated into distinct clusters. Centroids are placed in order to reduce the distance between measured points and clusters \cite{kmeans4}.

\begin{figure}[!h]
  \begin{center}
    \includegraphics[scale=0.48]{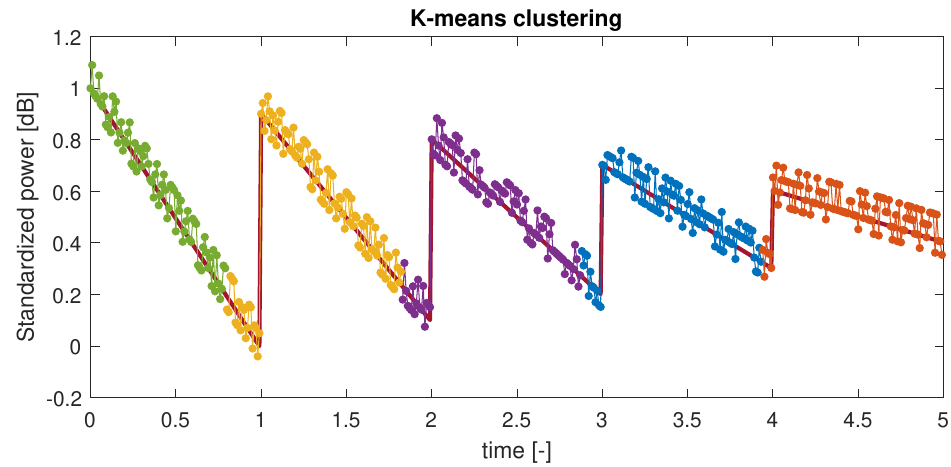}
  \end{center}
  \caption{Clustering based on k-means algorithm on synthetic signal}
  \label{fig:simul_kmeans}
\end{figure}

Simulations of the given algorithm are shown in Fig.~\ref{fig:simul_kmeans}. It is obvious that k-means clustering produces inaccurate results because includes the tail of the previous cluster in the new one.

\section{Sparsity based clustering- reweighted l1 minimization} \label{section_rwl}
The SV model is based on the assumption that multipath propagation  occurs in clusters with an exponential decrease in received power. Therefore, the proposed approach partitions the measured PDP into several segments, or distinct clusters. The concept of the algorithm proceeds in two steps \cite{rwl1}: 
\begin{itemize}
    \item reconstruction of transmit signal $\hat{P}$ from measured PDP.
    \item identification of the clusters from the shape of $\hat{P}$.
\end{itemize}
The following non-convex optimization problem is formulated from the aforementioned concept \cite{rwl2}:
\begin{align}
    \label{rwl:optimization}
    \min_{\hat{\boldsymbol{P}}} \norm{\boldsymbol{P}-\hat{\boldsymbol{P}}}_2^2+\lambda\norm{\Omega_2 \cdot \Omega_1 \cdot \hat{\boldsymbol{P}}}_0,
\end{align}
where $\boldsymbol{P} = [P(0),\ldots,P(N-1)]^T$ denotes the received signal and similarly for the reconstructed signal $\hat{\boldsymbol{P}}$, and $\lambda$ is a positive regularization parameter.
The feasible set $\mathcal{X}$ for the minimization problem \eqref{rwl:optimization} is
\begin{equation}
    \mathcal{X} = \{\hat{\boldsymbol{P}} \in \mathbb{R}_0^N\mid \norm{\Omega_2 \cdot \Omega_1 \cdot \hat{\boldsymbol{P}}}_1 \leq L_{\text{max}}\}.
    \label{mathcalX}
\end{equation}
The finite-difference operator  $\Omega_1$ is used for estimating the slope of $\hat{\boldsymbol{P}}$. The turning point, where the slope changes significantly, is determined using the finite-difference operator $\Omega_2$, represented by the filter [1, -1], defined in \cite{rwl1} as:

\begin{equation}
\Omega_2 = 
    \begin{bmatrix}
    1 & -1 & 0 & \hdots & \hdots & 0\\
    0 & 1 & -1 & \hdots & \hdots & 0\\
    \vdots & \ddots & \ddots & \ddots & \ddots & \vdots\\
    0 & 0 & \ddots & 1 & -1 & 0\\
    0 & 0 & \hdots & \hdots & 1 & -1\\
    \end{bmatrix}_{(N-2)\times(N-1)} .
\end{equation}

The reweighted $\ell_1$ norm minimization strategy \cite{rwl3} is used by \cite{rwl1} to numerically approximate a sparse solution and a more accurate clustering result.

The reweighted $\ell_1$ minimization \cite{rwl3} is applied to the optimization problem (\ref{rwl:optimization}) for iteratively increasing the sparsity of the solution by the weighted norm. The algorithm consists of the following phases \cite{rwl1}.
\begin{enumerate}
    \item Set the iteration count $m$ to zero and the initial weights $w_n^{(0)}=1$, \mbox{$n=1,...,N$.} In the reweighted $\ell_1$ minimization, the weight parameter is later employed to ensure a sparse solution.
    \item The weighted $\ell_1$ minimization problem is as follows:
    \begin{equation}
    \label{rwl:minimization}
    \begin{split}
    \hat{\boldsymbol{P}}^{(m)}= \arg \min_{\hat{\boldsymbol{P}}\in\mathbb{R}_0^N} \norm{P-\hat{\boldsymbol{P}}}_2 \hspace*{12ex}
    \\
    \text{s.t. } \norm{W^{m}\cdot\Omega_2 \cdot \Omega_1 \cdot \hat{\boldsymbol{P}}}_1 \leq L_{\text{max}}
    \end{split}
    \end{equation}
    where $L_{\text{max}}$ is the upper bound on the number of clusters in a PDP. $W^{(m)}$ is a matrix with weights at diagonal.
    \item The weights are updated according to
        \begin{equation}
    \label{rwl:W}
    w^{(m+1)}_n =\frac{1}{\hat{P}^{(m)}(n)+\epsilon},\quad (n=1,\ldots,N).
    \end{equation}
    In order to maintain stability and make sure that a zero-valued component in $\hat{\boldsymbol{P}}^{(m)}$ does not strictly preclude a nonzero estimate at the following phase, the constant $\epsilon$ is utilized. Epsilon can be selected as the smallest value permitted in the computing environment or a value significantly smaller than the anticipated nonzero magnitudes of $\boldsymbol{P}$. In our case for testing, the $\epsilon$ was equal to $10^{-9}$.
    \item When the method has run for the specified number of iterations $M$ or when the weights have converged, the algorithm terminates. If not, increment $m$ and move on to step 2.
\end{enumerate}

The recovered $\hat{\boldsymbol{P}}$ signal (purple color) for the simulation case is depicted in Fig. \ref{fig:synthetic}.

\begin{figure}[!h]
  \begin{center}
    \includegraphics[scale=0.5]{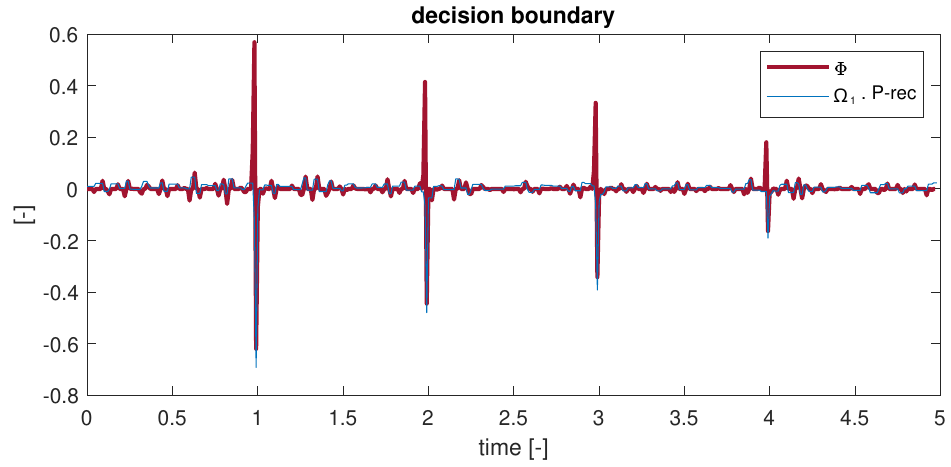}
  \end{center}
  \vspace{-15pt}
  \caption{Decision boundary for synthetic signal according to $\Phi$ by reweighted $\ell_1$ minimization}
  \label{fig:dec_bound}
\end{figure}

\begin{figure}[!h]
  \begin{center}
    \includegraphics[scale=0.5]{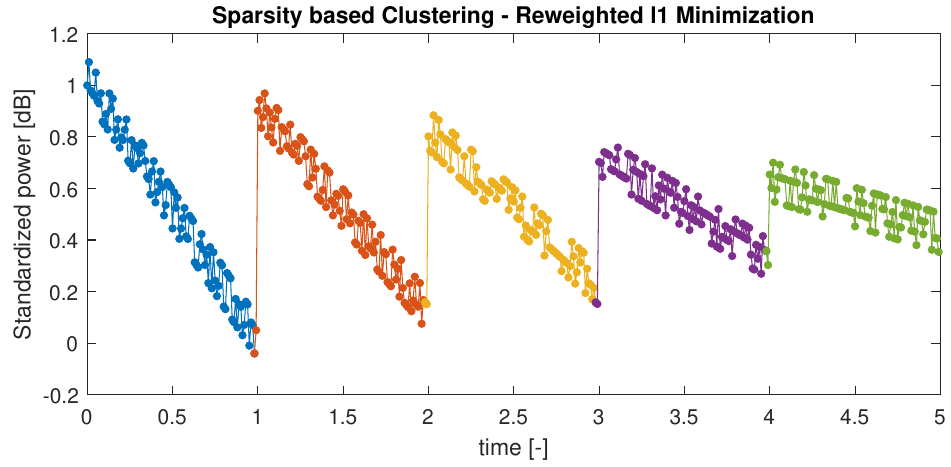}
  \end{center}
  \caption{Synthetic signal divided into clusters by based clustering algorithm with the use of reweighted $\ell_1$ minimization}
  \label{fig:clusters_by_rwl}
\end{figure}

It is feasible to locate the beginnings of clusters by using the recovered signal $\hat{\boldsymbol{P}}$, which is piecewise linear. The following vector
\begin{equation}
\label{phi:points}
\Phi=\left[\Omega_2 \cdot \Omega_1 \cdot \hat{\boldsymbol{P}}\right]_{(N-2)\times 1},
\end{equation}
denotes the pivot point of $\hat{\boldsymbol{P}}$ where the slope dramatically changes and can be used to pinpoint clusters with accuracy. In Fig. \ref{fig:clusters_by_rwl} there is a synthetic signal from Fig. \ref{fig:synthetic} divided into clusters automatically by presented sparsity based clustering algorithm which uses reweighted $\ell_1$ minimization according gained decision boundary shown in Fig. \ref{fig:dec_bound}


\section{Experimental measurements} \label{section_exp_me}
The measurements were taken as part of a measuring campaign \cite{vut_measure} to analyze in-vehicle communication and signal propagation inside car cabin.

The transmit and receive open-ended waveguide antennas are used for measurements inside a mid-sized vehicle Skoda Octavia~III. The RX antenna is positioned on the front windshield adjacent to the rear-view mirror. The TX antenna is positioned in the middle of the back seat. A complete description of the measurement campaign may be found in \cite{vut_measure}.


\subsection{55-65 GHz measurement apparatus}
In the 55-65 GHz frequency range, the transmission coefficient between two antennas is measured using the R$\&$S ZVA67 four-port VNA. A broadband power amplifier (QuinStar QPW-50662330, measured gain of 35 dB in the band of interest) was used on the transmitting side of the measurement setup to increase the system dynamic range. Antennas for transmitting and receiving signals were WR15 open waveguide, which has a radiation pattern identical to \cite{vut_measure} (similarly as in \cite{ant}). This measurement setup has a system dynamic range of about 50 dB. The IF bandwidth of 100 Hz and the VNA output power of 5 dBm were selected.

A 10 MHz frequency step is used during the measurement, which is carried out in the frequency domain. The inverse Fourier transform with Blackman windowing is used to convert the channel frequency response into the time domain. Phase-stable coaxial cables were utilized to prevent the TX antenna motions from reducing the observed phase accuracy. The power amplifier was calibrated together with the all four ports. In order to equalize the VNA transmission in the forward direction, the TX and RX antennas flanges were coupled together. The open-ended waveguide utilized for the experiment at 55-65 GHz has a radiation pattern that is 120° wide in the E-plane and 60° wide in the H-plane (for 3 dB decay).


\subsection{Application clustering methods on measured PDP based on \mbox{in-vehicle} communication}

A significant number of PDPs in a variety of scenarios were gathered during the measurement campaign outlined above. It is straightforward to show that the channel inside the car cabin and the measurement inside the room are quite comparable, allowing an SV model to be used. Therefore, in-vehicle PDP can be clustered using the suggested approaches.

\begin{figure}[!h]
  \begin{center}
    \includegraphics[scale=0.55]{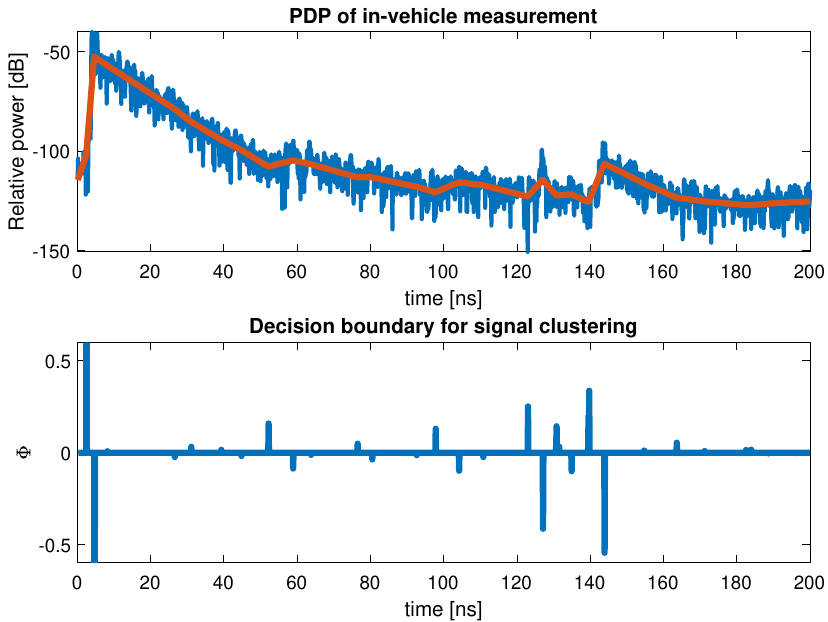}
  \end{center}
  \caption{PDP of measured in-vehicle communication (Scenario 1) (blue) and reconstructed $\hat{P}$ signal (red) using reweighted $\ell_1$ minimization and Decision boundary for clustering of measured signal}
  \label{fig:CIR21_measured}
\end{figure}

\begin{figure}[!h]
  \begin{center}
    \includegraphics[scale=0.55]{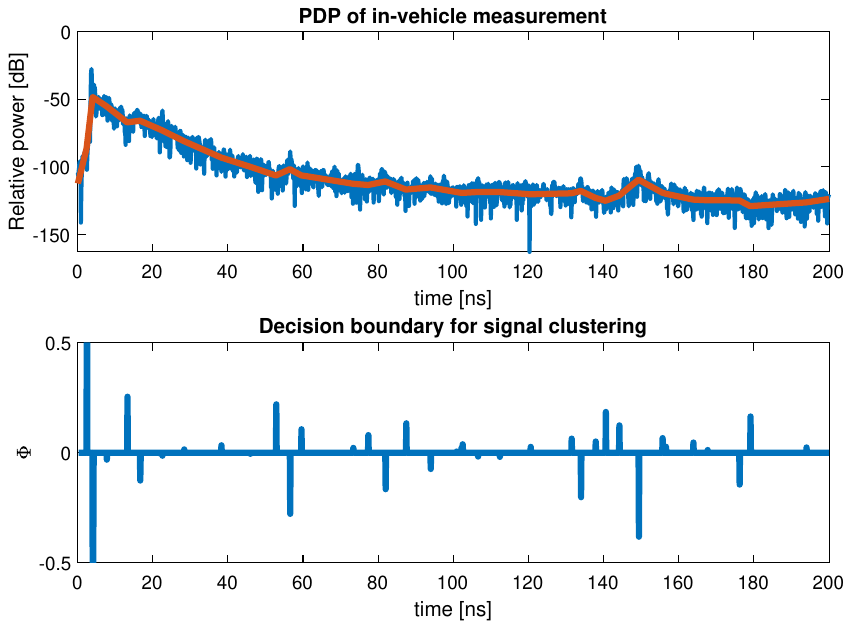}
  \end{center}
  \caption{PDP of measured in-vehicle communication (Scenario 2) (blue) and reconstructed $\hat{P}$ signal (red) using reweighted $\ell_1$ minimization and Decision boundary for clustering of measured signal}
  \label{fig:CIR12_measured}
\end{figure}

The illustrative PDPs from measurement configuration are shown in Fig. \ref{fig:CIR21_measured}, Scenario 1- the transceiver in the location behind the driver, the trunk door was open and car cabin without passengers and Fig. \ref{fig:CIR12_measured} Scenario 2- with open windows instead trunk door. The reconstructed signal $\hat{P}$ is shown with red color, and it is used to detect crucial turning points by using the sparsity-based clustering-reweighted $\ell_1$ minimization approach.

Decision boundary $\Phi$ according to (\ref{phi:points}) is shown in Fig.~\ref{fig:CIR21_measured} and \ref{fig:CIR12_measured} below, respectively. There is no denying that some pikes are noticeably higher than others. Once the decision level is set to -0.35, value optimized for this measurement scenario, clusters can be found and the PDP divided into clusters of MPC.

\begin{figure}[!h]
  \begin{center}
    \includegraphics[scale=0.53]{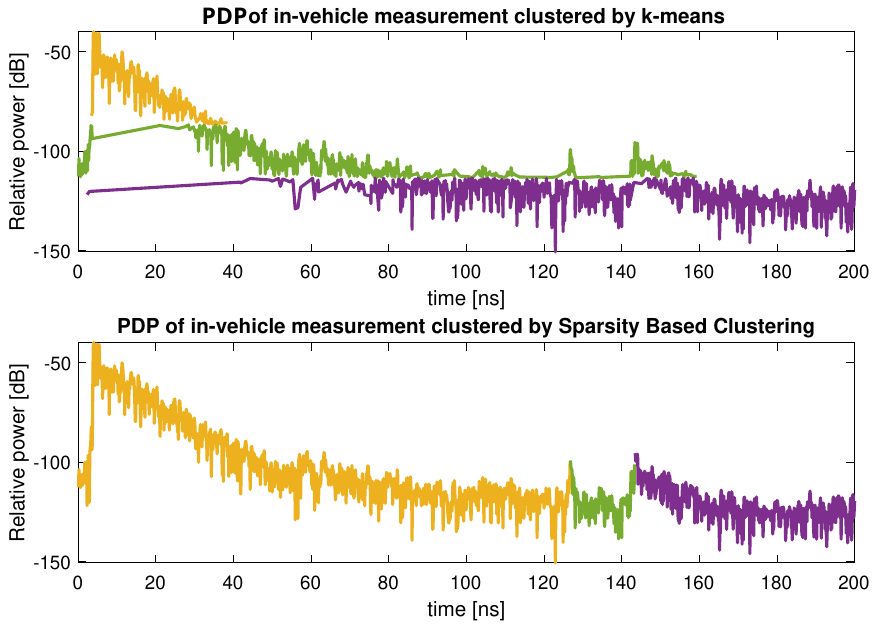}
  \end{center}
  \caption{PDP of in-vehicle measurement (Scenario 1) clustered by k-means algorithm and Sparsity Based Clustering- Reweighted $\ell_1$ minimitazion}
  \label{fig:CIR21kmeans_sparsity}
\end{figure}

\begin{figure}[!h]
  \begin{center}
    \includegraphics[scale=0.55]{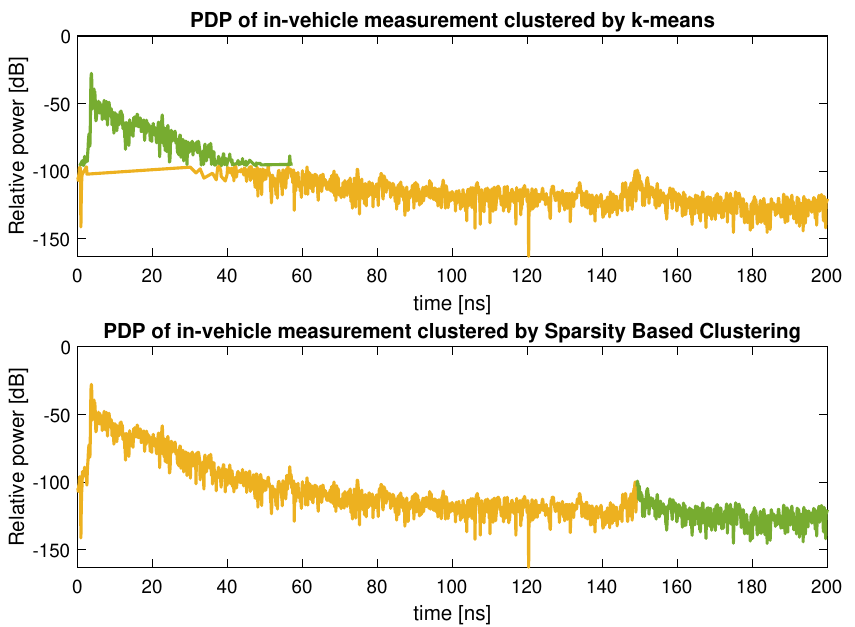}
  \end{center}
  \caption{PDP of in-vehicle measurement (Scenario 2) clustered by k-means algorithm and Sparsity Based Clustering- Reweighted $\ell_1$ minimitazion}
  \label{fig:CIR12kmeans_sparsity}
\end{figure}

PDPs clustered using k-means and sparsity-based clustering, respectively, are displayed in Figs. \ref{fig:CIR21kmeans_sparsity} and \ref{fig:CIR12kmeans_sparsity}. The common k-means approach is obviously inappropriate for time domain data since it divides PDP into a horizontal plane rather than separating clusters into a vertical plane. Compared to that, Sparsity based clustering segregated the signal with excellent precision in accordance with the decision boundary in Figs. \ref{fig:CIR21_measured} and \ref{fig:CIR12_measured} using Reweighted $\ell_1$ minimization.  

In scenario 1, points that indicate the beginning of new clusters have a markedly higher intensity than other points. This is due to the fact that the decrease in PDP is practically exponential (a straight line on a logarithmic scale) and exhibits minimal deviation. Additionally, the reconstructed signal for scenario 2 appears to be very smooth, but as can be seen in the graph of the decision boundary for scenario 2, the signal $\Phi$ contains more intense spikes, making it crucial to set the horizontal level for point selection very carefully.

\section{Conclusions}
We investigated clustering for partitioning MPCs into distinct segments. The clustering of the PDP is highly advantageous for estimating easily interpretable channel parameters. PDP data was collected during an extensive measurement campaign focused on in-vehicle communication at 60 GHz. The delay domain representation of the PDP was subjected to both k-means and sparsity-based clustering techniques. The k-means technique is less suited for analyzing time domain data, primarily due to its horizontal rather than vertical delay data clustering. However, the sparsity-based clustering approach proved highly effective for accurately segmenting real-world PDP data obtained from in-vehicle measurements. It was confirmed that the Saleh-Valenzuela model is adequate for in-vehicle communication at 60 GHz.

\section*{Acknowledgment}
This work was supported by the Czech Science Foundation, Lead Agency Project No.~22-04304L, Multi-band prediction of millimeter-wave propagation effects for dynamic and fixed scenarios in rugged time varying environments. The infrastructure of the SIX Center was used.

\bibliographystyle{IEEEtran}
{\footnotesize
\bibliography{IEEEabrv,biblio_yousef.bib}
}


\end{document}